\documentclass[11pt]{article}

\usepackage{graphicx}  
\begin{document}   
\bibliographystyle{plain}

\def\RR{\rm \hbox{I\kern-.2em\hbox{R}}}
\def\NN{\rm \hbox{I\kern-.2em\hbox{N}}}
\def\ZZ{\rm {{\rm Z}\kern-.28em{\rm Z}}}
\def\CC{\rm \hbox{C\kern -.5em {\raise .32ex \hbox{$\scriptscriptstyle
|$}}\kern
-.22em{\raise .6ex \hbox{$\scriptscriptstyle |$}}\kern .4em}}
\def\L{\pounds}
\def\<{\langle}
\def\>{\rangle}
\def\nl{\newline}
\def\vp{\varphi}
\def\e{\varepsilon}
\def\s{{\cal S}}
\def \vo {{\mathbf{\omega}}}
\def \ve {{\mathbf{\epsilon}}}
\def \mL {{\mathbf{\Lambda}}}
\def \mLt {{\mathbf{\Lambda}^t}}
\def \mLi {{\mathbf{\Lambda}^{-1}}}
\def \mS {{\mathbf{\Sigma}}}
\def \mSt {{\mathbf{\Sigma}^t}}
\def \mSi {{\mathbf{\Sigma}^{-1}}}
\def \Cov {{\mbox{Cov}}}
\def \Var {{\mbox{Var}}}

\def\b{\begin{equation}}
\def\e{\end{equation}}

\newtheorem{Tm}{Theorem}[subsection]
\newtheorem{Df}[Tm]{Definition}
\newtheorem{Lm}[Tm]{Lemma}
\newtheorem{Cr}[Tm]{Corollary}
\newcommand{\Rq}{\addtocounter{Tm}{1}{\bf Remark \theTm } \hspace{0.15cm}}
\newcommand{\Ex}{\addtocounter{Tm}{1}{\bf Example \theTm } \hspace{0.15cm}}
\newenvironment{Pf}{\begin{quote}{\bf Proof} \hspace{0.03cm}}{{$\Box $}
\newcommand{\ss}{\scriptscriptstyle}
\newcommand{\s}{\scriptstyle}
\end{quote}}

\begin{center}

{\Large   \bf A multivariate multifractal model for return fluctuations\\}
\vspace*{1cm}   
E.Bacry$^a$,  J. Delour$^b$, J.F. Muzy$^{b,c}$
\end{center}

\vspace*{0.6cm}  
\noindent 
$^a$ Centre de Math\'ematiques Appliqu\'ees, Ecole Polytechnique, \\ 91128 Palaiseau Cedex, France. \\
$^b$ Centre de Recherche Paul Pascal, Avenue Schweitzer,
 33600 Pessac, France. \\
$^c$ URA 2053, CNRS, Universit\'e de Corse, Grossetti, 20250 Cort\'e, France. \\

\vspace{1cm}

\begin{center}
{\large \bf Abstract}
\end{center}
\vspace{0.5cm}

\noindent

In this paper we briefly review the recently inrtroduced Multifractal Random Walk (MRW)
that is
able to reproduce most of recent empirical findings
concerning financial time-series :  no correlation between price variations,
long-range volatility correlations and multifractal statistics. We then focus on its extension to
a multivariate context in order to model portfolio behavior.
Empirical estimations on real data suggest that this approach can be pertinent to account for the nature of both linear and
non-linear correlation between stock returns at all time scales.

\section{Introduction}
Multifractal processes 
and the deeply connected mathematics of large deviations and
multiplicative cascades have been widely used in many contexts to account for the time scale dependence of
the statistical properties of a time-series. 
Recent empirical findings \cite{aArn98,pFis97,pBra99,pSch99}
suggest that in finance, this framework is likely to be pertinent.
The recently introduced Multifractal Random Walks (MRW) \cite{prl} are multifractal
processes that have proved successful to model return fluctuations. They can be seen as
simple ``stochastic volatility'' models (with stationary increments) whose statistical
properties can be precisely controlled across the time scales using very few parameters.
In that respect, they reproduce many features that characterize market price changes \cite{epj}
including the decorrelation of the price increments, the long-range correlation of the
volatility and the way the probability density function (pdf) of the price
increments changes across time-scales, going from quasi Gaussian distributions at rather
large time scales to fat tail distributions at fine scales. 

In a recent work \cite{pf}, Muzy et. al. have elaborated a ``multivariate multifractal''
framework  that accounts for the time scale dependence of
the mutual statistical properties of several time-series.
They have shown that the statistical properties of financial time-series can be described
within this framework. Though initially introduced for modelling single asset variations,  the MRW models can be naturally
extended in order to fit this new multivariate framework. Thus, the so-obtained Multivariate MRW (MMRW) can be used to
reproduce precisely the statistical properties of several assets at any time-scale. This is of course particularly useful for
modelling a portfolio behavior. 

The goal of
this paper is to explain how MMRW models are built and to
show, using real data, that eventhough they involve very few parameters,
they allow one to capture not only linear correlation between assets but also non linear correlation at all time scales.
The paper is organized as follows. In section 2, after recalling briefly the
main notations and definitions involved in the ``classical'' monovariate multifractal framework,
we introduce the MRW model defined in \cite{prl}, recall its main properties and show that it
reproduces precisely, at any time-scale, the statistical properties of real financial time-series. In
section 3, after presenting the multivariate multifractal framework defined in \cite{pf}, we
introduce the MMRW model and perform some analytical computations of its multifractal statistics.
Numerical estimations of the parameters in the case of financial times-series are extensively discussed.
Conclusions and prospects are reported in section 4.

\section{The Multifractal Random Walk (MRW) model}
\subsection{Multifractal processes and cascade models}
A multifractal process is a process wich has some scale invariance properties. These properties are generally
characterized by the exponents $\zeta_q$ which govern the power law scaling of the absolute moments of its
fluctuations, i.e.,
\b
\label{moment}
M(q,l) = K_q l^{\zeta_q},
\e
where
\[
M(q,l) = E\left(|\delta_l X(t)|^q\right) = E\left(|X(t+l)-X(t)|^q\right),
\]
where $X(t)$ is supposed to be a stochastisc process with stationary increments. 
Some very popular stochastic processes are the so-called {\em
self-similar processes} \cite{bTaqq}. They
are defined as processes $X(t)$ which have stationary increments and which
verify (in law)
\[
\delta_{l} X(t) =^{law}
(l/L)^H \delta_L X(t), ~~\forall l,L >0.
\]
For these processes, one easily gets 
$\zeta_q = qH$, i.e., the $\zeta_q$ spectrum is a linear function of $q$.  Widely used examples of such
processes are (fractional) Brownian motions (fBm) or Levy walks. 

However, many empirical studies have shown
that the $\zeta_q$ spectrum of return fluctuations is a non linear convex function. 
Let us note that, using a
simple argument, it is easy to show that if $\zeta_q$ is a non-linear convex function
the scaling behavior (\ref{moment}) cannot hold for all scales $l$ but only for scales smaller than an
arbitrary large scale $T$ that is generally referred to as the {\em integral scale}.
A very common approach originally proposed
by several authors in the field of fully developed turbulence
\cite{nov,sl,FP,dg,castaing}, has been to describe such
processes in the scale domain, describing the {\em cascading} process that rules
how the fluctuations evolves when
going from coarse to fine scales. Basically, it amounts in stating that the fluctuations at the integral scale
$T$ are linked to the ones at a smaller scale $l<T$ using the cascading rule
\b
\label{castaingd}
\delta_{l} X(t) =^{law} W_{l/T} \delta_T X(t)
\e
where $W_{l/T}$ is a log infinitely divisible stochastic variable which depends only on the ratio $l/T$.
A straightforward computation \cite{castaing} then shows that the pdf $P_l(\delta X)$ of
$\delta_l X$ changes when varying the time-scale $l$ according to the rule
\b
\label{castaing}
P_{l}(\delta X) = \int G_{l/T}(u) e^{-u} P_{T}(e^{-u} \delta
X) du,
\e
where the {\em self-similarity kernel} $G_{l/T}$ is the pdf of $\ln  W_{l/T}$. Since $W_{l/T}$ is a log
infinitely divisible variable, the Fourier transform of $G_{l/T}$ is of the form
\b
\label{G}
{\hat
G}_{l/T}(k) = {\hat
G}^{\ln l/T}(k).
\e
From
that equation, one easily gets the expression of the $\zeta_q$ spectrum
\b
\label{zz}
\zeta_q = \ln {\hat G}(-iq).
\e
Thus,  the simplest non-linear case is the so-called log-normal model that corresponds to a
parabolic
$\zeta_q$ and a Gaussian kernel.

Multiplicative cascading processes are examples of processes satisfying the cascading rule
(\ref{castaingd}). However, they have fundamental drawbacks: they do not lead
to 
stationary increments and they do not have continuous scale invariance
properties, i.e., Eq. (\ref{castaingd}) and consequently Eq (\ref{moment}) 
only holds for discrete scales $l_n
= \lambda^n$. To our knowledge, the MRW's are the only known multifractal  processes
with continuous dilation invariance
properties and stationary increments.

\subsection{Introducing the  MRW model}
An MRW process $X(t)$ is the limit process  (when the time discretization step $\Delta t$ goes to 0) of a
standard random walk $X_{\Delta t}[k]$ with a stochastic variance (volatility), i.e.,
$$
X(t) = \lim_{t\rightarrow 0} X_{\Delta t} (t),
$$
with
$$
X_{\Delta t} (t) = \sum_{k=1}^{t/\Delta t}
\epsilon_{\Delta t}[k] e^{\omega_{\Delta t}[k]},
$$
where $e^{\omega_{\Delta t}[k]}$ is the stochastic volatility and $\epsilon_{\Delta t}$  a
gaussian white noise of variance $\sigma^2 \Delta t$ and which is independant of $\omega_{\Delta t}$. The
choice for the process $\omega_{\Delta t}$ is simply dictated by the fact that we want the scaling
(\ref{moment}) to be exact for all time scales $l\le T$. Some long but straightforward computations
\cite{prl} show that this is achieved if $\omega_{\Delta t}$ is a stationary Gaussian process such that
$
E\left(\omega_{\Delta t}[k]\right) = -\Var\left(\omega_{\Delta t}[k]\right)$ and
whose covariance is
$$\Cov(\omega_{\Delta t}[k],\omega_{\Delta t}[l]) =
\lambda^2 \ln \rho_{\Delta t} [|k-l|]$$
where 
$$
\rho_{\Delta t} [k] =
\left\{
\begin{array}{ll}
\frac{T}{(|k|+1)\Delta
t} & \mbox{for}~|k|\le T/\Delta t -1
\\
1 & \mbox{otherwise}
\end{array}
\right.
$$
Let us note that it corresponds to a log-normal volatility which is correlated up to a time lag $T$.

One can then prove \cite{prl} the multifractal scaling property
\b
\label{moment2}
M(q,l) = K_q l^{\zeta_q},~~~\forall l \le T,
\e
with
\begin{equation}
 \label{zetamodel}
\zeta_q = (q-q(q-2)\lambda^2)/2.
\end{equation}
Since $\zeta_q$ is a parabolic function, it indicates that the self-similarity kernel $G_{l/T}$
which links the pdf's at different time scales (Eq. (\ref{castaing})) is Gaussian.
Moreover one can show \cite{prl} that the {\em magnitude} correlation $C_{\omega}(l,\tau)$
 defined
by
\b
  \label{cfmag1}
C_{\omega}(\Delta t,l) = Cov\left(\ln|\delta_l X(t)|,\ln|\delta_l X(t+\tau)|\right), 
\e
behaves like
\b
  \label{cfmag}
 C_{\omega}(\Delta t,l) 
 \sim -\lambda^2 \ln\left(\frac{\Delta t}{T}\right),~~~l<T  \; .
\e

\subsection{Modelling return fluctuations using MRW}
MRW processes can be used to model return fluctuations \cite{epj}. For this purpose 3 parameters need to be
estimated : the variance $\sigma$, the integral scale $T$ and the {\em intermittency parameter} $\lambda$
(it is called this way since it controls the non linearity of the $\zeta_q$ spectrum and consequently it
controls ``how much stochastic'' is the variable $W_{l/T}$). The variance $\sigma$ can be estimated using the
simple relation $Var(X(t)) = \sigma^2 t$. Both, the decorrelation scale $T$ and the parameter $\lambda$ can be
obtained from the expression (\ref{cfmag}) of the magnitude correlation. Let us note that $\lambda$ can be
also estimated independantly from the $\zeta_q$ spectrum (Eq. (\ref{zetamodel})). The consistency between these
two completly different estimators of $\lambda$ is a very good test for the validity of the model.

Parameter estimations have been made on financial data (japenese Yen futures). As shown in  figure 1, 
the MRW reproduces very precisely both the parabolic $\zeta_q$ spectrum (which  describes,  through Eqs (\ref{zz}),
(\ref{castaing}),  how the return fluctuation pdf evolves when going from one time scale to another) 
and the correlation structure of the magnitude. 

Let us remark that one can show
that $K_{q}=+\infty$ (in Eq. (\ref{moment2})) if $\zeta_q < 1$ and thus the pdf of
$\delta_l X(t)$ has fat tails \cite{prl}. In order to control the order of
the first divergent moment (without changing $\lambda$), one could simply
choose for
the $\epsilon_{\Delta t}$'s a law with fat tails (e.g. t-student laws).

\begin{figure}
  \begin{center}
    \includegraphics{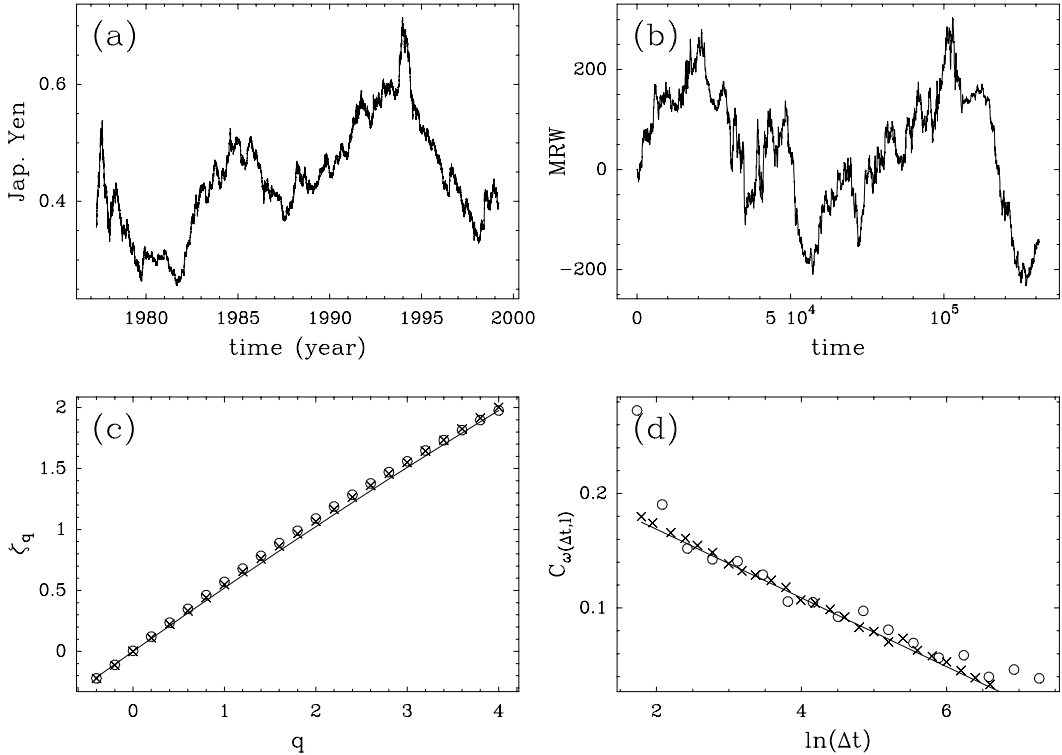}
    \caption{
      {\bf Modelling intraday Japenese Yen futures using MRW}.
The MRW is used to model the de-seasonalized logarithm of the
time-series displayed in (a). The parameters have been estimated to   
$\sigma^2= 4.10^{-6}$,
$\lambda^2 = 0.03$ and
$T = 1$ year.      
(a) Plot of the original index time-series : Japenese Yen futures from March 77 to February 99 (intraday tick by tick data). 
(b) Plot of a sample time series of length $2^{17}$ of the model.
(c) $\zeta_q$ spectrum estimations for the Yen futures
fluctuations ($\circ$) and for the MRW model ($\times$). The solid line corresponds to the
theoretical prediction (Eq. (\ref{zetamodel})). 
(d) Magnitude correlation function estimations as defined in Eq. (\ref{cfmag1})
(with $l\simeq 4$ days) for the Yen futures fluctuations ($\circ$) and for the MRW model ($\times$). The solid line corresponds to
the theoretical prediction (Eq. (\ref{cfmag})).}

    \label{fig_1}
  \end{center}
\end{figure}

\section{The multivariate multifractal random walk  (MMRW) model}
\subsection{The multivariate multifractal framework}
In this section, we generalize in a very natural way the multifractal framework introduced in section 2.1 to
multivariate processes. This generalization is inspired from \cite{pf}. It basically consists in
rewriting the cascading rule (\ref{castaingd}) using multivariate processes. Thus if ${\bf X} = (X_1,\ldots,X_N)$ is a
multivariate process, we will assume that it satisfies
\[
\left\{\delta_{l} X_i(t)\right\}_{1\le i \le N} =^{law} \left\{W_{i,l/T} \delta_T X_i(t)\right\}_{1\le
i \le N}
\]
where ${\bf W} = \left\{W_{i,l/T} \right\}_{1\le i \le N}$ is a log infinitely divisible stochastic vector
which depends only on the ratio
$l/T$. A straightforward computation \cite{castaing} then shows that the pdf $P_l(\delta {\bf X})$ of
$\delta_l {\bf X}$ changes when varying the time-scale $l$ according to the rule
\b
\label{mcastaing}
P_{l}(\delta {\bf X}) = \int du_1 \ldots \int du_N G_{l/T}({\bf u}) e^{-u_1-\ldots -u_N} P_{T}(e^{-{\bf u}} \otimes \delta
{\bf X}) ,
\e
where ${\bf u} = \{u_1,\ldots,u_N\}$ and
where the $G_{l/T}$ is the pdf of $\ln  {\bf W}_{l/T}$. (We used the notation 
${\bf a} \otimes {\bf b} = (a_1b_1,\ldots,a_N b_N)$). In the same way as in
section 2.1, one can easily get the scaling law of the moments
\b
\label{mmoment}
M(q_1,\ldots,q_N,l) = E \left( |\delta_l X_1(t)|^{q_1}\ldots |\delta_l X_N(t)|^{q_N}\right)
= 
K_{q_1,\ldots,q_N} l^{\zeta_{q_1,\ldots,q_N}},
\e
where the multifractal spectrum $\zeta_{q_1,\ldots,q_N}$ is linked to the self-similarity kernel through the relation
\[
\zeta_{q_1,\ldots,q_N} = {\hat G}(-iq_1,\ldots,-iq_N),
\]{*equation}
where $\hat G$ is defined as in Eq. (\ref{G}). 

\subsection{Introducing the MMRW model}
In order to account for the fluctuations of financial portfolios and
to consider management applications of our approach, it is important
to build a multivariate version of the MRW model. Since only Gaussian random variables are
involved in the construction of section 2.2,
this generalization can be done in a very natural way \cite{epj}.
The MMRW walk ${\bf X}(t)$ is defined as
$$
{\bf X}(t) = \lim_{t\rightarrow 0} {\bf X}_{\Delta t} (t),
 = \lim_{t\rightarrow 0} \sum_{k=1}^{t/\Delta t}
{\bf \epsilon}_{\Delta t}[k] \otimes e^{{\bf \omega}_{\Delta t}[k]}.
$$
(We again used the notation 
${\bf a} \otimes {\bf b} = (a_1b_1,\ldots,a_N b_N)$).
The process $\epsilon_{\Delta t}$ is Gaussian with zero mean and covariance $Cov(\epsilon_{i,\Delta t}(t),
\epsilon_{j,\Delta t}(t+\tau)) = \delta(\tau)\mS_{ij} \Delta t$. The matrix $\mS$ quantifies the variance and the correlation of
the different white noises involved in each component of ${\bf X}$.  We will refer to this matrix as the ``Markowitz matrix''. The
magnitude process ${\bf \omega}_{\Delta t}$ is Gaussian with covariance
$Cov(\omega_{i,\Delta t}(t),
\omega_{j,\Delta \tau}) = \mL_{ij} \ln(T_{ij}/(\Delta t + |\tau|))$ (for $\Delta t + |\tau| < T_{ij}$) and 0 elsewhere, where the
matrix
$\mL$  controls the non-linearity of the multifractal spectrum so we will refer to it as the ``multifractal matrix''. Moreover,
as in section 2.2, the mean of the process is chosen so that
$E({\bf
\omega_{\Delta t}}) = -Var({\bf
\omega_{\Delta t}})$. Let us note that the previously defined coefficients $\sigma^2$ and
$\lambda^2$ for an asset $i$ correspond respectively to
the diagonal elements $\mS_{ii}$ and $\mL_{ii}$.

In order to show that MMRW are multivariate multifractal processes (within the framework of the previous section),
we would like now to compute the $\zeta_{q_1,\ldots,q_N}$ spectrum. There are 2 cases for which this computation is basically
the same as for the regular MRW model : (i) the case were all the white noises are decorrelated (i.e., $\mS$ is diagonal), (ii)
the case where the stochastic variances of all the assets correspond to the same process, i.e.,
$\omega_{i,\Delta t} =
\omega_{j,\Delta t}, ~\forall i,j$. In both cases, a straightforward computation shows that the scaling law (\ref{mmoment})
holds $\forall l \le \min_{ij}(T_{ij})$ and the spectrum is
\b
\label{mzeta}
\zeta_{q_1,\ldots,q_N} = \sum_{i=1}^{N} \zeta^i_{q_i} - \sum_{1\le i<j\le N} \mL_{ij} q_i q_j,
\e
where $\zeta^{i}_{q}$ refers to the spectrum of the $X_i$ component of ${\bf X}$. The computation of the spectrum is trickier in
the general case. However one can show that, in this case, all the extra terms (compared to the particular case (i)) that
appear in the development of
$M(q_1,\ldots,q_N,l)$ go to 0 when $\Delta t \rightarrow 0$ \cite{inprep}.
Consequently, the multifractal spectrum has the same expression.

Since the spectrum is a parabolic function, it indicates that the self-similarity kernel $G_{l/T}$
which links the pdf's at different time scales (Eq. (\ref{mcastaing})) is Gaussian.
Moreover one can show \cite{inprep} that the magnitude correlation behaves like
\b
  \label{mcfmag}
Cov\left(\ln|\delta_\tau X_i(t)|,\ln|\delta_\tau X_j(t+l)|\right)
 \sim -\mL_{ij} \ln(l)+C  \; .
\e

\subsection{Modelling return fluctuations using MMRW}
For modelling a basket of assets using an MMRW, one needs to estimate the Markowitz matrix $\mS$, the multifractal matrix $\mL$
and the different integral scales $T_{ij}$. As for the MRW, the magnitude correlation can be used for estimating both the
integral scales and the multifractal matrix. The Markowitz matrix $\mS$ can be estimated, after having
estimated $\mL$, by using the simple relation
\begin{equation}
\label{covx}
   \Cov(X_{i}(l),X_{j}(l)) =
   \mS_{ij}e^{\frac{1}{2}(\mL_{ii}+\mL_{jj}+2\mL_{ij})}l
   \; .
\end{equation}
Let us note that $\mL$ can be
also estimated independantly from the $\zeta_{q_1,\ldots,q_N}$ spectrum (Eq. (\ref{mzeta})). 
Indeed, to estimate $\mL_{ij}$ one could simply estimate, for instance, the exponent of the power law scaling
\begin{equation}
\label{rr}
   R_{ij,q}(l) = \frac{E(|X_i(l)|^q|X_j(l)|^q)}{E(|X_i(l)|^q)E(|X_j(l)|^q)}
\sim
l^{-\mL_{ij}q^2}
\end{equation}
The consistency between these
two  different estimators of $\mL$ is a very good test for the validity of the model.

Parameter estimations have been made on some assets (daily data) of the cac40. Figure 2 shows 
the estimations of the non diagonal term $\mL_{12}$ of the multifractal matrix
$\mL$ using both the magnitude correlation estimator (Eq. (\ref{mcfmag}))  and the $R_{12,q}(\Delta t)$ estimator (Eq.
(\ref{rr})). These two different estimators lead to very close  estimations of $\mL_{12}$  which is consistent with
the MMRW model.

On Figure 3 we have displayed the histograms of all the non diagonal terms $\mL_{ij}$ (resp. $\mS_{ij}$ and $T_{ij}$) for all
the pair of assets in the cac40. Eventhough the histogram of the $\mL_{ij}$ is pretty wide, its maximum is reached for
$\mL_{ij} \simeq 0.02$ which is the most common value found when estimating $\mL_{ii} =
\lambda_i^2$ in the monovariate case \cite{epj}. In the same way the histogram of the $T_{ij}$ (resp. $\mS_{ij}$) has a peak
around 1-2 years (resp. $4. 10^{-6}$) which also corresponds to the most common value found when estimating $T_{i}$ (resp.
$\sigma^2$) in the monovariate case
\cite{epj}. These results suggest that, as a first approximation, one could model these assets using an MMRW which shares the
same magnitude process for all the assets, i.e.,
$\omega_i =
\omega_j~\forall~Êi,j$. Though it is clearly not exactly the case, it simplifies the model a lot and allows to perform many
analytic computations.

\begin{figure}
  \begin{center}
    \includegraphics{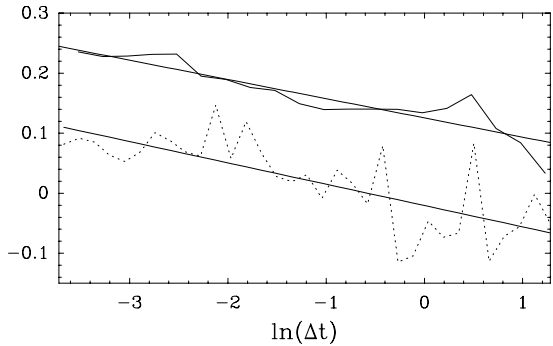}
    \caption{
      {\bf Estimation on real data of a non diagonal term $\mL_{ij}$ of the multifractal matrix $\mL$.} $\mL_{12}$ is estimated
using daily data (from 1993 to 2000) where the first asset is AXA and the second one is TOTAL-FINA. As explained in the
text, two estimators can be used. The solid curve corresponds to the $R_{12,q=1}$ estimator (Eq. (\ref{rr}))  which leads to the
estimation $\mL_{12} \simeq 0.032$. The dashed curve corresponds to the covariance estimator (Eq. (\ref{mcfmag}))
 which leads to the estimation $\mL_{12}\simeq 0.036$. These two estimators lead to consistent estimation of $\mL_{12}$. }
\end{center}
\end{figure}

\begin{figure}
  \begin{center}
    \includegraphics{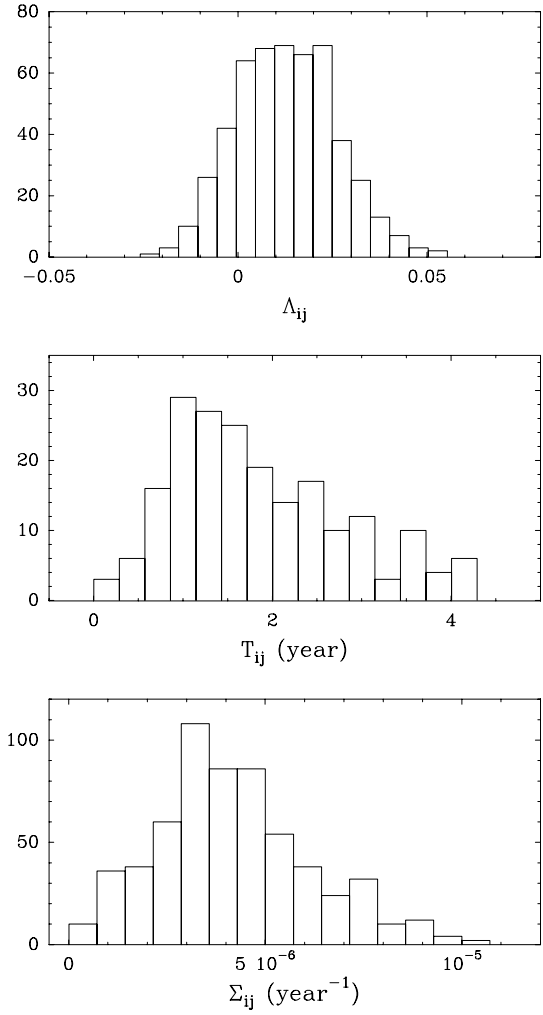}
    \caption{
      {\bf Parameters estimations for a basket made of all the cac40 assets.}
Histogram of the estimations, for all
the pair of assets in the cac40, of the non diagonal terms (a) $\mL_{ij}$, (b)
$T_{ij}$ and (c)
$\mS_{ij}$ (using Eq. \ref{rr}).
}
    \label{fig_3}
  \end{center}
\end{figure}

\section{Conclusion}
In this paper, we have introduced a multivariate model for return fluctuations that is a definite step beyond standard
correlation analysis. It  basically corresponds to a multivariate random walk using a stochastic volatility.
This model can be characterized using the recently introduced notion of multifractal multivariate that itself relies on the idea
that the simplest way to describe the statistics of a process at all time scales is to assume some scale invariance properties.
Consequently, the MMRW has a potentiality to capture the whole return joint law of a basket of assets at all time horizons.
As
shown in this paper, we are able to reproduce the main observed characteristics of financial
time-series: no correlation between price variations, long-range volatility correlations, linear and non-linear
correlation between assets and the price increment pdf and the way it changes when varying the time-scale. All of these features
can be controlled using a few parameters : the multifractal matrix which controls the scale invrariance properties, the integral
scales which controls the volatility correlation and the Markowitz matrix which controls the noise correlation. Moreover, as we
have already pointed out, in good approximation, one can consider that all the assets share
the same volatility process. Not only it reduces the number of parameters but it makes any analytical computation much easier. 
We are currently applying MMRW for portfolio management and risk control.

\section{Acknowledgement}
  We acknowledge Matt Lee and Didier Sornette for the
  permission to use their financial data.
We are also very grateful to Alain Arneodo 
 and Didier Sornette for interesting discussions.
\noindent
All the computations in this paper have been made using
  the free GNU licensed sofware {\em LastWave}~\cite{lw}.

\end{document}